\documentclass[12pt]{article}
\DeclareUnicodeCharacter{2061}{}
\usepackage{amsmath,amssymb}
\usepackage{graphicx}
\usepackage{float}

\usepackage{scicite}

\usepackage{times}

\newenvironment{sciabstract}{%
\begin{quote} \bf}
{\end{quote}}

\newcounter{lastnote}

\title{Building and aligning a 10-plane light converter}
\author
{Ohad Lib$^{1}$, Ronen Shekel$^{1}$, and Yaron Bromberg$^{1\ast}$\\
\\
\normalsize{$^{1}$ Racah Institute of Physics, The Hebrew University of Jerusalem, Jerusalem 91904, Israel}\\
\\
\normalsize{$^\ast$To whom correspondence should be addressed; E-mail: Yaron.Bromberg@mail.huji.ac.il.}
}

\date{}

\begin{document} 

\maketitle

\begin{sciabstract}
The ability to manipulate the spatial structure of light is fundamental for a range of applications, from classical communication to quantum information processing. Multi-plane light conversion (MPLC) addresses the limitations of single-plane modulation by enabling full control over multiple spatial modes of light through a series of phase masks separated by free-space propagation. In this tutorial, we present a step-by-step guide for building and aligning a 10-plane programmable light converter using a single spatial light modulator (SLM) and standard optical components. Our method allows precise alignment, achieving single-pixel accuracy with a relatively simple setup. We hope that this guide will help other researchers to quickly adopt and adapt MPLC technology for their own experiments.
\end{sciabstract}

\section*{Introduction}

The encoding of information in the spatial structure of light stands at the basis of various applications, ranging from classical communication\cite{puttnam2021space} and imaging\cite{bertolotti2022imaging} to quantum information processing\cite{erhard2020advances,lib2022quantum}. To enable these applications and harness the potential of such spatial modes of light, methods to efficiently manipulate and control them are required. This is typically achieved by using spatial light modulators (SLMs) that can arbitrarily control the phase or amplitude of a given spatial mode. However, to obtain full control and apply arbitrary transformations on multiple free-space spatial modes, a single SLM is not sufficient, as it only applies phase or amplitude patterns onto the incoming light without interfering or overlapping spatially separated areas of the beam. 

By going beyond a single plane of light modulation, multi-plane light conversion (MPLC) overcomes this limitation and enables full control over multiple spatial modes of light\cite{morizur2010programmable}. To this end, a series of phase masks separated by free-space propagation are used, where any desired transformation is encoded by appropriately optimizing the different phase masks, and diffraction enables the mixing of spatially separated areas of the beam. Since its first demonstration over a decade ago, MPLC has been used for various applications with classical light\cite{labroille2014efficient,fontaine2019laguerre,kupianskyi2023high, butaite2022build, kupianskyi2024all, martinez2024reconfigurable, ran2024enhancing}. More recently, this technology has been adopted for quantum information processing, where the spatial modes of single\cite{brandt2020high} and entangled \cite{lib2022processing} photons are manipulated for applications in quantum interference, computation\cite{lib2024resource}, and communication experiments\cite{goel2023simultaneously, lib2024high}. A similar multi-plane approach has also been adopted in the context of diffractive neural networks\cite{lin2018all, zhou2021large, yu2025all}.

In this tutorial, we detail how to build and align a 10-plane programmable light converter (fig.\ref{fig:1}a,b). While fixed MPLCs with pre-designed phase masks have been thoroughly explored and are even commercially available for certain applications\cite{barre2017broadband}, programmable MPLCs having a large number of planes have only been recently demonstrated\cite{lib2024resource, rocha2025self}. Here, we first discuss how to build a 10-plane programmable MPLC using a single commercial SLM and simple optical components. We then describe a procedure that uses the programmability of the MPLC to perform a precise, yet relatively simple, alignment of the positions of the MPLC phase masks with respect to the manipulated light. As MPLCs can be built using common optical components and a single SLM, and algorithms for calculating the desired phase masks are already publicly available\cite{fontaine2019laguerre}, we hope our detailed guide will assist interested groups in quickly adopting this exciting technology.

\section*{Building a 10-plane programmable light converter}

Before building and aligning an MPLC, one should conduct simulations to find the optimal number of planes, their desired locations on the SLM, the free-space propagation distance, and the number of pixels in each phase mask required for their specific application. To do so, one could start from an educated initial guess for the different parameters. For example, the distance between planes could be set such that the different modes start to diffract significantly (e.g. a few Rayleigh ranges for a Gaussian beam), the number of planes can be set to be larger than the number of modes if arbitrary transformations are desired, and the number of pixels per mode in each phase mask can be set to achieve decent sampling. Once a set of initial parameters have been determined and a target transformation has been identified, corresponding phase masks can be calculated via wavefront-matching \cite{fontaine2019laguerre} or gradient descent \cite{kupianskyi2023high} methods. The fidelity of the transformation can then be optimized by tweaking the different parameters, choosing the ones to be implemented in the experiment.

To be concrete, we discuss the build and alignment procedure with our specific parameters and design, as the principles are universal and should be helpful regardless of the particular parameters chosen.

\begin{figure}[H]
\centering
\includegraphics[width=1\textwidth]{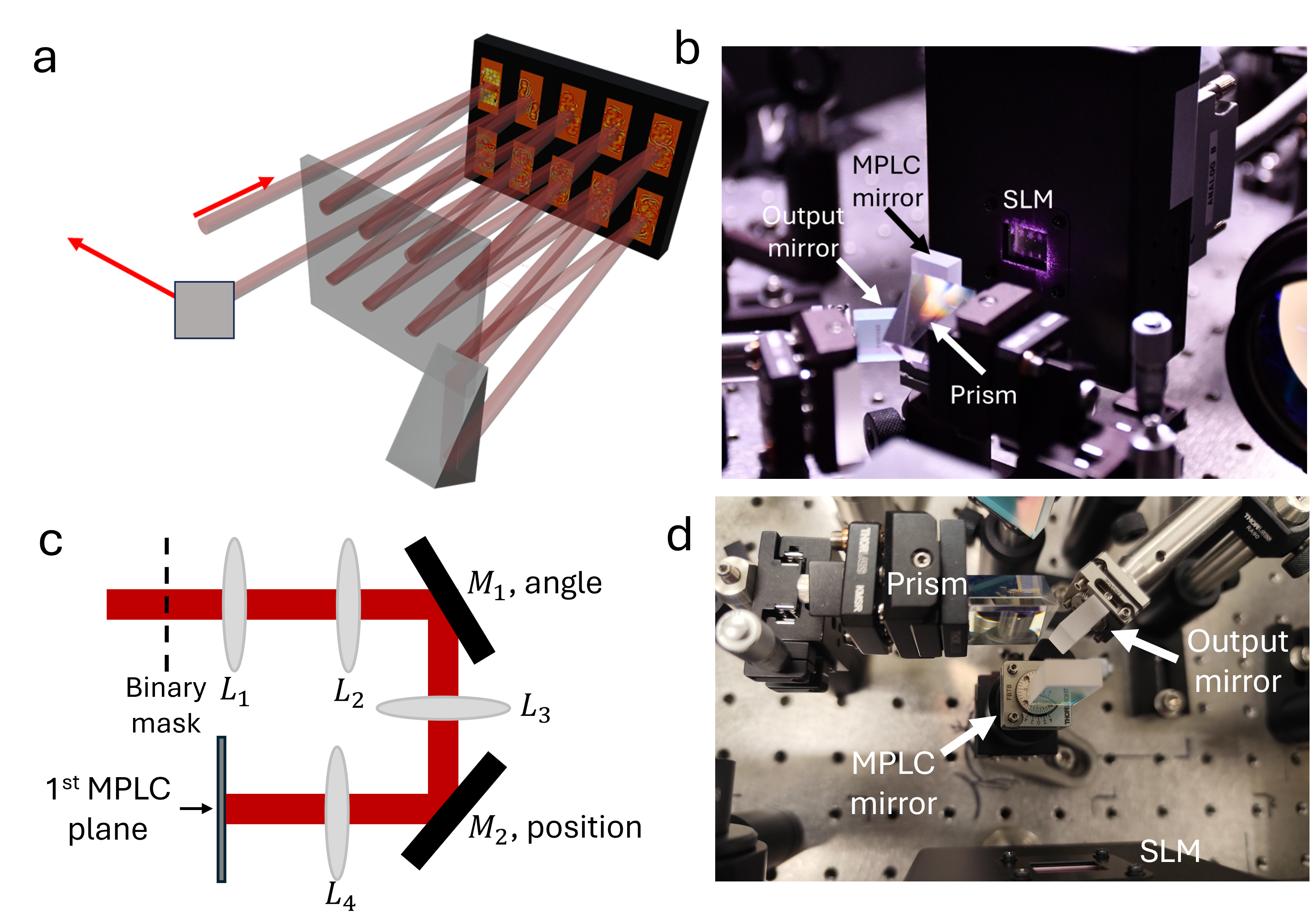}
\caption{\label{fig:1} \textbf{10-plane light converter}. (a) 10 phase masks separated by free-space propagation are implemented using a single SLM, a mirror, and a prism. The light bounces between the SLM and the MPLC mirror five times on the top part and then five times on the bottom part. (b), (d) Side and top view pictures of the experimental implementation showing the SLM, prism, MPLC mirror, and output mirror. (c) A sketch of the first part of the experimental setup. M1 and M2 are mirrors, L1-L4 are lenses. In our experiment, a binary amplitude mask defines spatially separated modes that are imaged to the first plane of the MPLC\cite{lib2024resource}. Two mirrors, one imaged onto the MPLC and one in the Fourier plane, are used to independently control the input angle and position of the light, respectively.}
\end{figure}

Our 10-plane light converter consists of three main components (fig.\ref{fig:1}b): a programmable phase-only SLM (Hamamatsu X13138-02), a dielectric mirror (Thorlabs, BBSQ1-E03) and a prism (Thorlabs, PS908H-B). In our case, the size of each phase mask is $140 \times 360$ pixels, and that of the SLM is $1272 \times 1024$ pixels. Therefore, we fit our phase masks on the SLM in two rows: the light reflects between the SLM and the mirror five times at the top part, reflects back using a prism, and then reflects between the SLM and the mirror five additional times at the bottom part of the SLM (fig.\ref{fig:1}a).

To build the MPLC, we first mount the SLM directly to the optical table using a 1-inch diameter post. We found that avoiding any degree of freedom in the mounting of the SLM is crucial to the stability of the MPLC. As the SLM is fixed, we use two mirrors placed at the image and Fourier planes of the first phase mask of the MPLC to control the input angle and position of the light, respectively (fig.\ref{fig:1}c). The MPLC mirror is positioned in front of the SLM at a distance of $43.5mm$, which we found suitable for our specific optical modes and transformations through numerical simulations. We trim the mirror to a width of $12.7mm$, to prevent clipping of the input and output light on either side of the MPLC. The mirror is then glued to a compact kinematic mount, allowing fine control over its different angles (Thorlabs, FBTB). Finally, a compact translation stage is used to align the mirror in front of the SLM (Thorlabs, LX10).

A prism is used to access the bottom part of the SLM after five reflections. The side of the prism is glued to a metal plate and mounted onto a kinematic mount (Thorlabs, KMSR). The height of the prism is controlled using a translation stage (Thorlabs, LX10). We place the prism so that no light impinges onto its center to avoid the $\approx 250\mu m$-wide region of poor optical quality. Due to mechanical constraints, the prism is placed at a distance of $69mm$ from the SLM, which is considered when calculating the phase masks using wavefront-matching, by setting the distance between planes 5 and 6 to $2\cdot69mm$. Still, we found using a prism superior to other retro-reflection solutions due to relatively minimal induced aberrations and polarization rotation. After the last five phase masks, at the output of the MPLC, a square 1/2" mirror is used to redirect the light toward the rest of the experimental setup.

\section*{Alignment procedure}

After discussing the components and construction of the MPLC, we can now describe its alignment. While only three main elements are involved, the number of planes and the sensitivity of the MPLC to phase mask positioning errors, on the order of a few pixels, make precise alignment crucial and challenging. Luckily, as we will now discuss, the programmability of the SLM itself helps us to set a reliable and easy-to-implement procedure for aligning the MPLC \cite{chen2017precise}. This procedure is general and could work for an arbitrary number of planes. 

The alignment consists of three main stages: (a) rough alignment of the first five planes to get the light through the first part of the MPLC, (b) rough alignment of the last five planes to get the light to the output, and (c) fine alignment of the entire MPLC. At the end of this procedure, the centers of the light beam at the different planes and the phase mask on the SLM will coincide such that the transformations in the experiment should reproduce those designed in the simulation. In our case, the light source used for the alignment and experiments with the MPLC are spatially entangled photons generated via Spontaneous Parametric Down-Conversion (SPDC) \cite{walborn2010spatial, schneeloch2016introduction}. Nonetheless, since the alignment relies on intensity measurements with a camera rather than correlations between single-photon detectors, the SPDC source effectively behaves as a spatially incoherent source. Indeed, the procedure below is general and can also be performed with classical light (ideally, linearly polarized for our polarization-sensitive liquid-crystal SLM \cite{lazarev2019beyond, yang2023review}).
 
\subsection*{Rough alignment}
Starting with stage (a), place the SLM at the image plane of mirror M1 (fig.\ref{fig:1}c), as can be verified, for example, by changing the angle of the mirror without changing the position of the light on the SLM. Ensure the light hits the SLM at roughly the desired location and input angle (by looking at the angle of the reflected light). After mounting the SLM, correction to the input angle and position can be made using mirrors M1 and M2, respectively.

Now, place the MPLC mirror in front of the SLM at the desired distance. From our experience, the distance does not yet have to be precise. It will be estimated later, and the precise distance could be considered when calculating the phase masks. Using the translation stage, move the mirror horizontally to avoid clipping the input beam. The light hitting the SLM at several positions should now be visible to the eye (or IR viewer). Change the angle of the MPLC mirror to make the distance between planes on the SLM (positions of the light hitting the SLM) look as equal as possible in both the vertical and horizontal directions. While more precise methods using an additional laser and a specialized mirror that is polished on both sides exist\cite{zhang2023multi}, we found this simple approach sufficiently accurate and much simpler. Then, use M1 and M2 to change the input position and angle to have five planes at roughly the desired positions at the top part of the SLM. If necessary, move the MPLC mirror in the horizontal direction to ensure the light is not clipped at the output.

Now that the light successfully passes through the first five planes of the MPLC, add the prism. Position it at the desired distance or as closely as possible to the SLM-mirror spacing. Change the vertical position of the prism to control the vertical position of the bottom five planes of the MPLC. Change the angle of the prism to roughly ensure that the horizontal positions of the fifth and sixth planes are similar. At this point, the bottom and top planes should all be roughly at the same horizontal positions, and the light should exit the MPLC at the entrance side, at the bottom part. If this rough alignment of the total ten planes is proven to be challenging in your setup, consider first performing the fine alignment to the first five planes (as will be described below) before proceeding to add the prism and the last five planes.

\subsection*{Fine alignment}

\begin{figure}[H]
\centering
\includegraphics[width=1\textwidth]{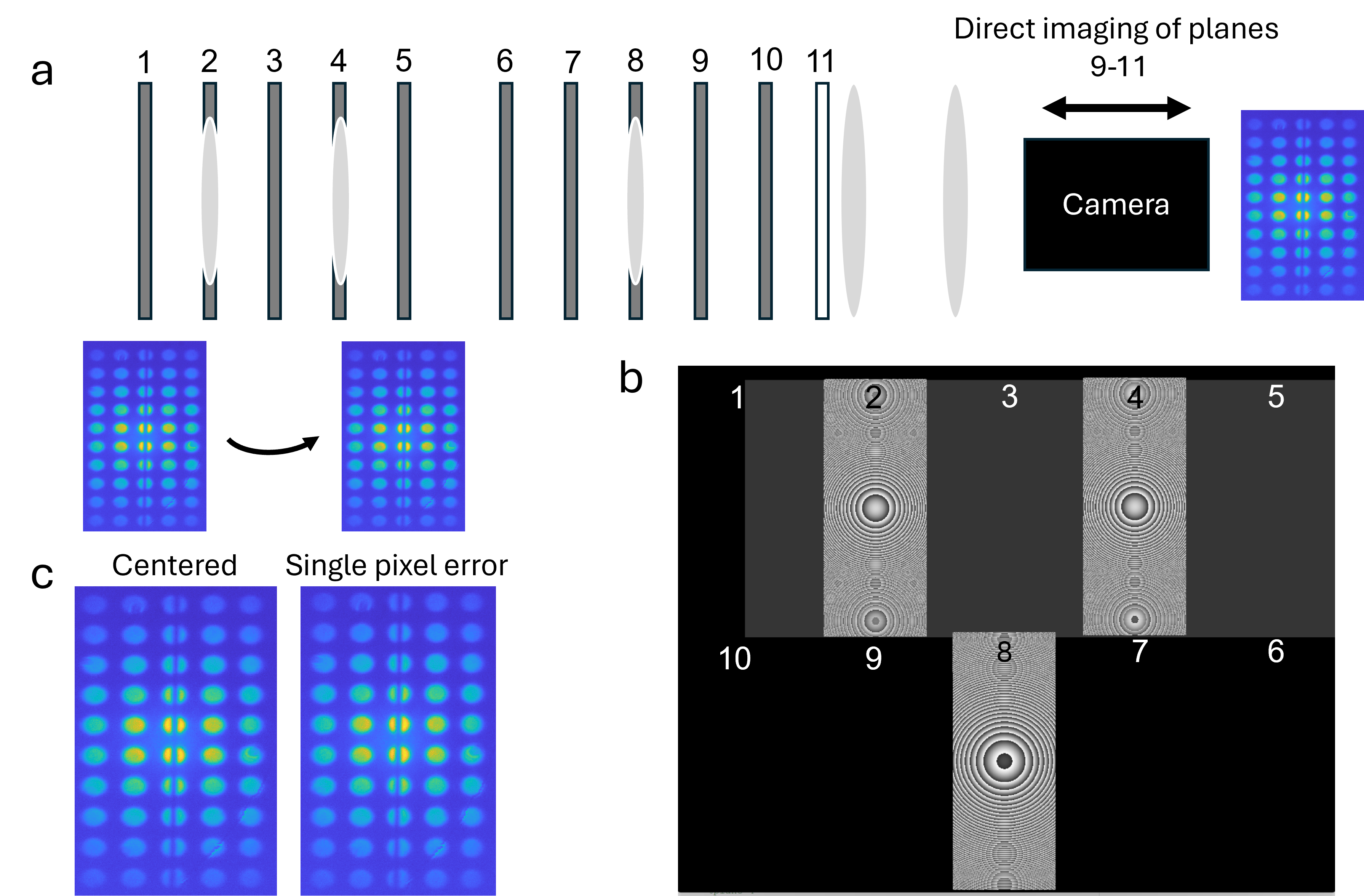}
\caption{\label{fig:2} \textbf{Fine alignment of the MPLC}. (a,b) The position of the first plane of the MPLC is found by imaging it onto the camera using effective lenses on planes 2, 4 and 8, and a $\pi$-step on plane 1. A correction pattern (not shown) is added to the SLM during the alignment process to compensate for its aberrations. (c) Single SLM-pixel accuracy is achieved by centering a $\pi$-step with the center of the beam. In our case, the light source consists of entangled photons passing through an array of 50 circular apertures, yielding the shape seen in the picture. The $\pi$-step on plane 1 is centered with the beam when it is centered with the middle column of light spots. On the right, upon moving the $\pi$-step by one pixel, clear asymmetry can already be seen on the spots at the center.}
\end{figure}

After the light passes through the entire MPLC, finer alignment must be performed to find the exact position where each phase mask should be placed on the SLM. This is a crucial step, as one must ensure that the center of the beam hits the center of the simulated phase mask in each plane. For this, add a camera at the output of the MPLC and use appropriate imaging lenses so that it can be translated to directly image the last three planes (Fig. \ref{fig:2}a). In our case, we calculate the MPLC transformations so that the output modes are located at plane 'eleven' which is positioned $2\cdot43.5 mm$ after the last physical MPLC plane, such that the propagation distance between planes ten and eleven is equal to twice the SLM-mirror spacing. Thus, we set the camera and imaging lenses to image planes 9, 10, and 11. 

During the alignment procedure, it is crucial for the center of the beam on the first plane to be well-defined. In our case, we use an array of 50 spatial modes in a $5 \times 10$ configuration, making it easy to define the center of the beam accurately. Alternatively, one can use a pinhole or any other suitable feature of the beam. 

The first step of the fine alignment is to find where the center of the beam hits the SLM on the first plane. Therefore, we need to be able to image the first plane of the MPLC to the camera. For this, we use the MPLC itself: add quadratic phase masks that implement Fresnel lenses on appropriate planes of the MPLC to image the first plane of the MPLC to planes 9-11, which the camera can directly image. For example, the first plane can be imaged onto the eleventh plane by adding lenses at planes 2, 4, and 8 (Fig.\ref{fig:2}a,b). The lenses at planes 2 and 4, each with a focal length that is equal to the spacing between planes, implement a 4f system between planes 1 and 5, which is then imaged to plane 11 using the single lens at plane 8 with a focal length $f$ such that $\frac{1}{u}+\frac{1}{v}=\frac{1}{f}$ holds for distance $u$ to plane 5 and $v$ to plane 11. At this point we do not yet know the exact distance between the SLM and the MPLC mirror, so the focal lengths of the lenses are only approximate. For our purpose at this point it is sufficient that they reasonably image the sharp features of the incident beam onto the camera. 

Once the first plane is imaged to the camera, add a $\pi$-step along the horizontal/vertical direction to find the y/x coordinate of the center of the first phase mask on the SLM. The $\pi$-step will be visible on the camera, and will look like a sharp dark line. Move the position of the $\pi$-step until it is centered with the beam, indicating the exact pixel on the SLM around which the phase mask should be centered (fig.\ref{fig:2}c). Note that the chosen center of the lenses on planes 2, 4, and 8 could only affect the transverse position of the image on the camera, and not the relative position of the $\pi$-step and center of the beam on plane 1. If the center of the beam is not in the desired location on the SLM, fix it now with mirror M2 using feedback from the camera: put the $\pi$-step at the desired x/y location and tilt mirror M2 until the center of the beam reaches it.

The alignment of the second plane is carried out in a similar manner by imaging it to the camera using other planes and using $\pi$-steps. The accuracy here will largely depend on the ability to precisely define the center of the beam in this plane. This highly depends on the specific modes used in the experimental application, which we assume are the same ones used for alignment. If the center of the beam is hard to define in this plane, the already-aligned first plane of the MPLC can be used to add a phase mask (e.g., a Fresnel lens) to change the beam shape on the second plane and make its center easier to define. If the second plane is not in the desired location, tilt mirror M1 to correct for this, and verify that the position of the first plane has not changed.

From the third plane onwards, diffraction makes it challenging to define the center of the beam precisely. To overcome this problem, use quadratic phase masks again, but this time to image the first plane (where the center of the beam is well defined) to the current plane (whose center you wish to find). This is done sequentially, starting from the third plane onwards, as it is now crucial to know the center of previous planes so that the Fresnel lenses that image the first planes to the current plane are positioned accurately. For example, after finding the position of planes 1 and 2, use plane 2 to image plane 1 onto plane 3. Use $\pi$-steps on plane 3 to find the center of the beam, then image plane 3 onto the camera using other planes (of which the position does not have to be precisely known yet). In this example, an error in the position of the lens in plane 2 will linearly propagate to an error in the center of plane 3. This is avoided by first finding the centers of the first and second planes, then proceeding to the third plane, etc.

After finding the positions of planes 1 to 5, one can easily detect and fix misalignments in the MPLC, manifested, for example, by uneven spacing between the planes. Take the measured x/y positions of the planes and fit them to a quadratic function $a_0 + a_1(n-1)+a_2(n-1)^2$, where $n$ is the plane number, beginning with $n=1$. The first term, $a_0$, is determined by the beam's position at the first plane, which was already set to the correct position using mirror M2. The second term, $a_1(n-1)$ is a linear term dependent on the input angle onto the SLM, which was also already corrected using mirror M1. The last term, which should ideally be zero, results from an undesired angle between the SLM and the MPLC mirror, which causes the angle of the light to change at each reflection, yielding a quadratic term under the paraxial approximation. To fix this, set the positions of the planes according to $a_0 + a_1(n-1)$, assuming the last term is zero. Create a 4f system between planes 1 and 5 using quadratic phase masks placed at planes 2 and 4. Add a $\pi$-step at plane 5 at the expected position $a_0 + a_1(5-1)$, and tilt the MPLC mirror until it coincides with the center of the beam. Once the MPLC mirror as well as the input angle and position of the light are aligned, find the positions of the centers of planes 1 to 5 once more to verify the alignment.

Before proceeding to find the positions of the last five planes, use the first five planes to estimate the spacing between the MPLC mirror and the SLM. For that, image the first plane to the fifth plane again using a 4f system with planes two and four. Add a $\pi$-step in both planes, and change the effective focal length of the lenses until reaching ideal imaging between the two planes, such that both $\pi$-steps are sharp. The focal length of the lenses corresponds to the distance traveled by the light between planes. In our system, we found this method gives an estimate with an accuracy of less than $1~mm$, where such an error did not affect the transformations we performed with the MPLC in a significant manner. The accuracy of this method was estimated by finding the focal length where clear defocus of the $\pi$-step is observed. This was then verified in experiments with phase masks calculated for different distances between planes, yielding the same optimal distance up to less than $1~mm$.

Proceeding to the bottom part of the MPLC, image the first plane onto the sixth plane, as well as the sixth plane onto the camera. Find the center of plane 6, as discussed above, and use the prism to make any corrections to the location. Now, continue in a similar manner to find the positions of planes 6 to 10. The distance between planes 5 and 6 can be estimated by imaging the first and the sixth planes onto the camera and finding the optimal focal length of the relevant lenses. The same method could similarly be extended sequentially to an arbitrary number of planes, which may be located on multiple SLMs.

Finally, once the alignment is completed, we further correct for aberrations in the optical setup. Adding the SLM's correction pattern (provided by Hamamatsu in our case) during the alignment improves its accuracy. We then further correct other aberrations after the alignment process by empirically adding spatially-varying (in our case, mode-dependent) phases at the first plane of the MPLC. In addition, to find the optimal axial position of the detectors (in our case, single-photon detectors), which should be placed at plane eleven, we use a camera as well. We image once more a $\pi$-step from one of the MPLC planes to plane eleven using Fresnel lenses and find the position where the $\pi$-step is the sharpest using a camera. We place our detectors in this plane.

\section*{MPLC performance}
The performance of a given MPLC depends on many factors, from the desired application, number of planes, and quality of alignment to pixel crosstalk and laser bandwidth. While some of these parameters are highly application-specific and can vary between experiments, we use numerical simulations to provide some intuition for the dependence of the performance on the number of planes and quality of alignment.

\begin{figure}[H]
\centering
\includegraphics[width=1\textwidth]{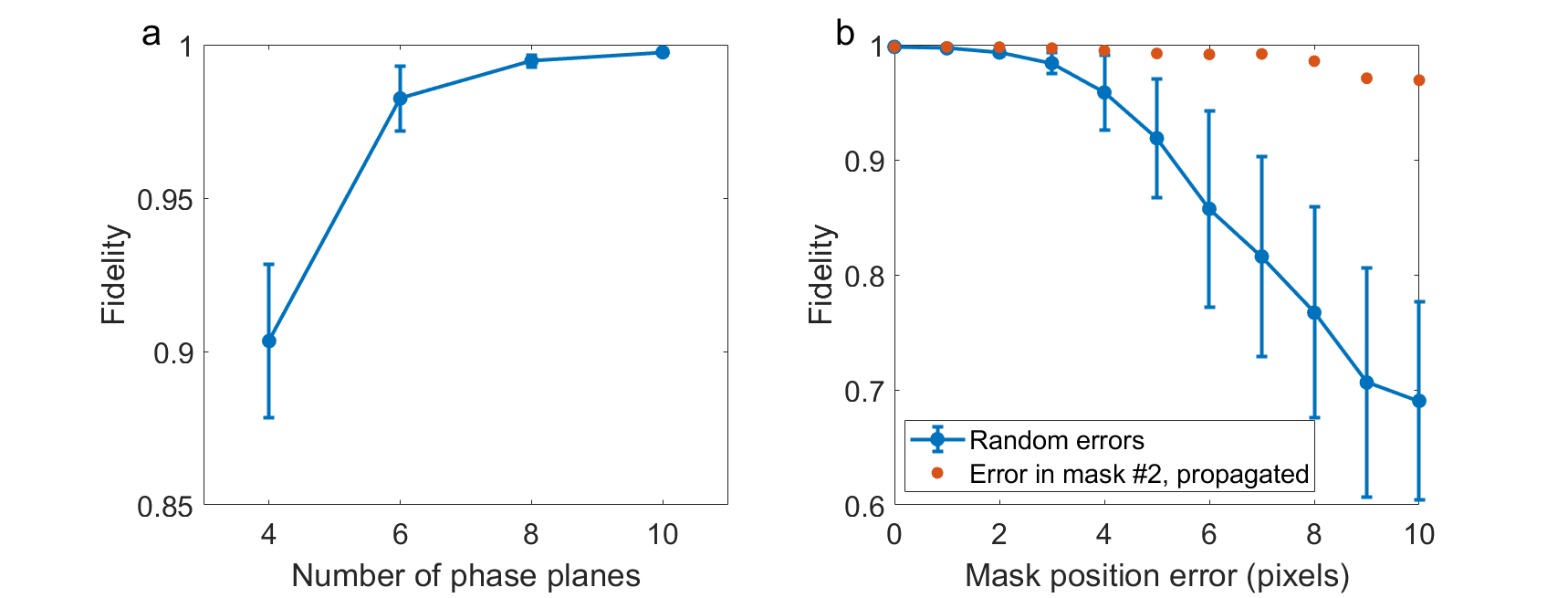}
\caption{\label{fig:3} \textbf{MPLC scaling and the effect of errors}. (a) The fidelity of a $5\times5$ random unitary transformation implemented using a different number of phase planes with equal spacings of $87 mm$. The error bars represent the standard deviation for the fidelity from different random unitaries. (b) For 10 phase planes, the fidelity of realizing a 5-dimensional discrete Fourier transform is presented as a function of position errors (in the horizontal direction) of the center of the masks. For random position errors, the x-axis marks the maximal position error for each mask, assuming uniform distribution of errors up to this number. For propagated errors from the second plane, the x-axis is the actual error applied to the second plane, which is then linearly propagated to the following planes. The effect of random position errors in each plane is stronger than that of a propagated error from an error in the second plane. All fidelities are calculated according to the Frobenius norm and are presented after correcting mode-dependent phases at the input and output planes.}
\end{figure}

Preliminary studies into the required number of MPLC planes as a function of the number of spatial modes have shown a linear scaling\cite{lopez2021arbitrary}, as one would expect from a naive counting of degrees of freedom. Furthermore, increasing the number of planes has been shown to increase the spectral bandwidth of the device and reduce the effect of pixel crosstalk\cite{fontaine2019laguerre}. It is therefore generally advisable to increase the number of MPLC planes, as long as the device's loss remains acceptable. We show a similar trend here by looking at the simulated fidelity of random unitary transformations on five Gaussian spatial modes as a function of the number of programmable phase planes (fig. \ref{fig:3}a). While the specific values of the fidelity could change significantly when taking into account different SLM-mirror spacings, pixel crosstalk strengths, and other imperfections, the trend of improved performance with an increasing number of planes is representative. This motivated our choice of building a large MPLC that takes advantage of the entire size of our SLM.

Increasing the number of planes can, however, induce larger alignment errors, which can lower the fidelity of the programmed transformations. From our alignment procedure described above, we can identify two leading types of errors in the positioning of the phase masks. The first is simply a random error in the position of each mask, resulting from the finite accuracy of finding the center of the beam on each plane. As shown above, this error is at the single-pixel level in our case. Another error is a systematic one, resulting from a random error in the position of one plane propagating to the following ones. This type of error is not corrected by our alignment procedure, which uses previous planes for the alignment of a given plane. The extreme case is an error in the position of the second plane, which will linearly propagate to all following planes. For example, a one-pixel error in the second plane means that the third plane will be aligned using a Fresnel lens with an offset of one-pixel, causing a two-pixel error.

Thankfully, as one can see in Fig. \ref{fig:3}b, errors propagating between planes, even when amounting to large errors on the last planes of the MPLC, do not significantly degrade the fidelity compared with random errors in each plane, motivating our choice of alignment method. We note that our approach is similar to the alignment of any optical setup. Each element is positioned as precisely as possible relative to the beam impinging on it, minimizing random errors in the position of each element. Errors can, however, propagate between optical elements. A shifted lens at the beginning of the setup can cause large beam deviations at the end of it, but the performance of the setup is not significantly degraded as long as all other components are aligned with respect to the shifted beam.

\section*{Conclusion}

In this tutorial, we hope to provide a detailed guide to building and aligning a 10-plane programmable light converter. While the initial alignment protocol is rather long, the final result is quite stable. In our setup, routine alignment is performed roughly once a month, which takes approximately an hour to complete. This is because, typically, only the transverse positions of the planes have to be found and slightly updated without physical changes to the mirrors or prism. The fact that this routine alignment requires physical movement only of the external camera and displaying phase masks on the computer-controlled SLM makes it easy to perform. Therefore, it is relatively simple to maintain the single-pixel accuracy of the MPLC. We hope that other groups in the community could use this tutorial to readily adopt and improve our proposed design and alignment procedure of a 10-plane light converter.

\section*{Acknowledgments}
This research was supported by the Zuckerman STEM Leadership Program and the Israel Science Foundation (grant No. 2497/21). O.L. acknowledges the support of the Clore Scholars Programme of the Clore Israel Foundation. R.S. acknowledges the support of the Center for Nanoscience and Nanotechnology, the Hebrew University, Israel; the Ministry of Innovation, Science and Technology, Israel; and the Council for Higher Education of Israel.

\bibliography{scibib}
\bibliographystyle{unsrt}

\end{document}